\documentclass[oneside,reqno,12pt]{amsart}

\usepackage{rangecite}
\usepackage{epsfig}

\numberwithin{equation}{section}

\newcommand{\lm}{\lambda}
\newcommand{\rh}{\rho}
\newcommand{\sg}{\sigma}
\newcommand{\bg}{\overline g}
\newcommand{\bR}{\overline R}
\newcommand{\bnabla}{\overline \nabla}
\newcommand{\tg}{\tilde g}
\newcommand{\tR}{\tilde R}
\newcommand{\tnabla}{\tilde \nabla}
\newcommand{\tG}{\tilde G}

\newcommand{\tpi}{\tilde \pi}
\newcommand{\tlmk}{{\tilde \lambda}_{k+1}}

\begin{document}

\vspace*{-0.577in}
\begin{flushleft}
\scriptsize{HEP-TH/9509147, UPR-661T}
\end{flushleft}

\vspace*{0.47in}

\title{Non-Trivial Vacua in Higher-Derivative Gravitation}
\author{Ahmed Hindawi, Burt A. Ovrut, and Daniel Waldram}
\thanks{Published in Physical Review D \textbf{53} (1996), 5597--5608}
\maketitle
\vspace*{-0.3in}
\begin{center}
\small{\textit{Department of Physics, University of Pennsylvania}} \\
\small{\textit{Philadelphia, PA 19104-6396, USA}}
\end{center}

\begin{abstract}

A discussion of an extended class of higher-derivative classical theories 
of gravity is presented. A procedure is given for exhibiting the new 
propagating degrees of freedom, at the full non-linear level, by transforming 
the higher-derivative action to a canonical second-order form. 
For general fourth-order theories, described by actions which are general 
functions of the scalar curvature, the Ricci tensor and the full Riemann 
tensor, it is shown that the higher-derivative theories may have multiple 
stable vacua. The vacua are shown to be, in general, non-trivial, 
corresponding to deSitter or anti-deSitter solutions of the original theory. 
It is also shown that around any vacuum the elementary excitations remain 
the massless graviton, a massive scalar field and a massive ghost-like 
spin-two field. The analysis is extended to actions which are arbitrary 
functions of terms of the form $\nabla^{2k}R$, and it is shown that such
theories also have a non-trivial vacuum structure.

\end{abstract}

\thispagestyle{empty}

\renewcommand{\baselinestretch}{1.2} \large{} \normalsize{}

\vspace*{\baselineskip}

\section{Introduction}

In a previous paper \cite{PRD-53-5583}, we presented a method for reducing 
a general quadratic theory of gravitation to a canonical second-order
form. The quadratic action provides an example of a higher-derivative
theory, where the gravitational equations of motion are higher than
second order. This means that the theory contains more degrees of
freedom than just the simple massless graviton.  By rewriting the
action, we showed that quadratic gravitation is classically
equivalent to Einstein's gravity coupled to a massive real scalar
field and a massive symmetric tensor field describing a spin-two
field, with a specific Lagrangian. 

We found that, even for simple quadratic gravity, the reduced action
gave highly non-trivial potential energy and kinetic coupling terms.
This suggests that higher-derivative theories have a great deal of
structure, not immediately apparent from a simple linear analysis
where one expands around flat space. One particular feature, which is
the main thesis of this paper, is that the transformed theory may have
a complicated vacuum structure. Here, by a vacuum solution, we mean a
stable solution of the second-order theory. In this paper, we will
only consider vacua in which the auxiliary fields are covariantly
constant. It is important to note that only once the vacuum of the
transformed theory has been identified can the nature of the
elementary field excitations be discussed. Thus, for instance, the
exact mass and couplings of the excitations, as well as which
excitations are potentially ghost-like, may be very different around
different vacuum states. In general, they will bear little relation to
the excitations of the linearized analysis. 
 
The vacuum structure of the quadratic theories remained comparatively 
simple. The only stable vacuum was flat space with zero vacuum expectation 
value for both of the auxiliary fields. The elementary excitations were a 
massive scalar field and a massive ghost-like spin-two field. We would like  
to investigate how other higher-derivative theories can introduce a more 
interesting vacuum structure. A natural generalization is to consider 
actions which are not quadratic but general functions of the curvature 
tensors. Since these tensors involve only 
second derivatives of the metric, the corresponding field equations can be at 
most fourth-order; that is, the same order as the quadratic actions. 
We can start by asking how many new degrees of freedom we expect in such 
theories. To give a rough count, we turn to the Cauchy problem. First 
recall that even higher-derivative gravity theories remain diffeomorphism 
invariant. Thus we are always able to transform away four components of the 
symmetric metric, leaving six free components. Since the field equations are 
fourth-order, if the Cauchy problem can be solved we expect to be required to 
give four initial conditions for each independent component of the metric, 
namely the component itself and the first three time derivatives. This 
would imply that we have at most twelve degrees of freedom. Rewriting the 
theory in a second-order form, since the auxiliary fields are set equal 
to terms involving second derivatives of the metric, we expect that 
giving the initial conditions of the auxiliary fields is equivalent to fixing 
second and third derivatives of the metric, a total of twelve conditions. 
Thus we find that the auxiliary fields should carry six degrees of freedom, 
just as in the general quadratic theory. This leaves a possible six further 
degrees of freedom in the metric. However, since the metric in the second-order 
theory obeys Einstein gravity, the number of degrees of freedom is 
reduced to the two helicity states of the usual massless graviton. By this 
rough argument, we expect an action in the form of any general function of 
the curvature tensors to describe, at most, the propagation of a massless 
graviton plus an additional six degrees of freedom; that is, the same as for 
the special case of the general quadratic theory discussed in our previous 
paper \cite{PRD-53-5583}. Be this as it may, the structure of such theories is 
potentially much richer than in the quadratic case. As we will see, the 
vacuum structure is now non-trivial. 

A second generalization is to consider gravity actions which 
have higher derivatives acting on the curvature tensors. Since these theories 
involve more than two derivatives on the metric, the corresponding field 
equations will be generically higher than fourth-order. Consequently, 
we now expect new degrees of freedom in addition to those of the general 
fourth-order theory. Again, we will show that the vacuum structure of such 
theories is, in general, non-trivial.

Some discussion of reducing both types of generalized theories to a
second-order form already exists in the literature. That the equations
of motion following from actions given by a general function $f(R)$
are equivalent to the equations of motion of a scalar field
dilatonically coupled to gravity, was first shown by Teyssandier and
Tourrenc~\cite{JMP-24-2793}. The equivalence at the level of
the action was given by Magnano 
\textit{et al.}~\cite{GRG-19-465,CQG-5-L95,CQG-7-557}, who
also rewrote the reduce theory in canonical form. This latter group
and Jakubiec and Kijowski~\cite{PRD-37-1406} also rewrote
actions given by a general function $f(R_{\mu\nu})$ in second-order
form, though without the canonical separation of the new degrees of
freedom. Actions including
derivatives of the scalar curvature were considered in the context of
inflation by Gottl\"ober \textit{et al.}~\cite{CQG-7-893}, who
again showed, in some special cases, the equivalence of the field
equations to those describing scalar fields coupled to gravity. Still
at the level of the field equations, this work was extended to a
general class of actions by Schmidt~\cite{CQG-7-1023} and
Wands~\cite{CQG-11-269}. Possible vacua of the $f(R)$ theory are
discussed, in terms of the original higher-derivative field equations,
by Barrow and Ottewill~\cite{JPA-16-2757}, and later in terms of
the second-order field equations by Barrow and
Cotsakis~\cite{PLB-214-515}. 

This paper is organized as follows. In the next three section we will 
consider actions which are general functions first of the 
curvature scalar only, then 
of the Ricci tensor, and finally of the full Riemann tensor. Rewriting these 
theories in a canonical second-order form, we will find a rich vacuum
structure.  This will require a careful discussion of the different
``branches'' of the theory, a concept we define below. We 
will show that the vacuum states are in a one-to-one correspondence with the 
stable, constant curvature, deSitter or anti-deSitter solutions of the 
higher-derivative theory. Furthermore, we will show that 
if the vacuum of the transformed theory has zero cosmological constant, then 
the corresponding spacetime for the original higher-derivative 
theory is also flat. We will then discuss the elementary excitations around a 
given vacuum and argue that they remain a graviton, a scalar field and 
a ghost-like spin-two field, with masses and couplings depending on the 
particular vacuum and the functional form of the original action. In Section 
\ref{fboxR} we will consider a class of gravity actions that are general 
functions of the curvature scalar $R$ and the derivatives $\nabla^{2k}R$, with 
$k$ a positive integer. We show how to rewrite such theories in second-order 
form and show that they have, in general, a non-trivial vacuum structure. In 
particular, we present an example with two new scalar degrees of freedom, 
neither of which is ghost-like, coupled to Einstein gravity with a stable 
anti-deSitter space as its vacuum. We briefly present our conclusions in 
Section \ref{conclusion}.

Throughout the paper our conventions are to use a metric of signature 
$(-+++)$ and define the Ricci tensor as
$R_{\mu\nu}=\partial_\lm\Gamma^\lm_{\ \mu\nu}
-\partial_\mu\Gamma^\lm_{\ \lm\nu}
+\Gamma^\lm_{\ \lm\rh}\Gamma^\rh_{\ \mu\nu}
-\Gamma^\lm_{\ \rh\mu}\Gamma^\rh_{\ \lm\nu}$. 

\section{Actions Given by General Functions of the Scalar Curvature}
\label{sec:f(R)}

As in the case of quadratic gravity, the higher-derivative theory
described by actions which are general functions of the curvature
scalar is classically completely equivalent to the canonical
second-order theory of a scalar field coupled to gravity. This
equivalence was first shown at the level of the action by Magnano
\textit{et al.}~\cite{GRG-19-465,CQG-5-L95,CQG-7-557}. We shall derive this 
equivalent
theory in a slightly different form, by first introducing an auxiliary
field to reduce the action to second order and then making a suitable
conformal transformation. We will point out the need to consider
different branches of the theory when making the reduction. 

We introduce the auxiliary field in two steps. First we write
\begin{align}
   S &= \frac{1}{2\kappa^2} \int{ d^4x \sqrt{-g} f(R) } \notag \\
     &= \frac{1}{2\kappa^2} \int{ d^4x \sqrt{-g} \left[ 
            f'(X) \left(R - X\right) + f(X) \right] },
\label{f(R)S}
\end{align}
where $f'(X)=df/dX$. The auxiliary field $X$ has the equation of motion
\begin{equation}
   f''(X) \left( R - X \right) = 0. 
\end{equation}
Provided we have $f''(X)\neq0$, this gives $X=R$. Substituting back into the 
action we return to the original higher-derivative form. 
Thus, the reduced action is equivalent to the original theory, but only away
from the critical points defined by $f''(X)=0$. A continuous region of $X$
between critical points, where $f''(X)$ never vanishes, will be called a
branch. Typically, there will be several branches in the theory. Since $X$ 
gets set equal to $R$, in terms of the original theory, the condition 
for a critical point can be written $f''(R)=0$. Thus, branches in $X$ 
correspond to branches in $R$ in the original higher-derivative theory.

As a concrete example, let us assume that
\begin{equation}
\label{eg1}
    f(R)=R+\epsilon^{-2} R^3,
\end{equation}
where $\epsilon>0$. Solving $f''(X)=0$ implies that there is a single critical
point at $X=0$. Therefore, the second-order formalism in terms of the 
auxiliary field $X$ is valid in each of the two branches $-\infty<X<0$ and
$0<X<\infty$, but breaks down at the critical point $X=0$. Since $X$ gets set
equal to $R$, these regions correspond to branches in the space of curvature
given by $\infty<R<0$ and $0<R<\infty$ with a critical point at $R=0$.

The second step, which will then allow us to rewrite the action in canonical 
form, is to change variables to a new field $\lambda=f'(X)$. In terms
of this new variable, the action becomes
\begin{equation}
\label{auxf(R)S}
    S =  \frac{1}{2\kappa^2} \int{ d^4x \sqrt{-g} \left[ 
         \lm \left(R - X(\lm)\right) + f(X(\lm)) \right] }.
\end{equation}
Clearly this action is well defined only for those regions where the $X$
action is valid; that is, away from the critical points. Therefore, the
$\lm$ formulation is only defined over ranges of $X$ for which
$f''(X)\neq0$. Note that to define the action \eqref{auxf(R)S} we must be able 
to invert the expression $\lm=f'(X)$ to give $X=X(\lm)$. Locally this 
requires the same non-degenerate condition $f''(X)\neq0$. Globally, there may 
still be many different roots when we solve for $X$ in terms of $\lm$. 
However, in any given branch of $X$ there is only a single root. Thus there 
is a valid formulation of the theory in terms of $\lm$ for each branch of 
$X$. It is important to note that, in each branch,  the inverted function 
$X=X(\lambda)$ which enters \eqref{auxf(R)S} is different. 

As a concrete example, consider once
again $f(R)$ defined in \eqref{eg1}. It follows that
\begin{equation}
\label{lmeg1}
    \lambda = 1 + 3 \epsilon^{-2} X^2.
\end{equation}
Note that $\lambda>1$. This expression can be inverted to give
\begin{equation}
\label{Xeg1}
    X= \pm \frac{\epsilon}{\sqrt{3}} \sqrt{\lambda-1}.
\end{equation}
Clearly one must take the $+$ root in the $0<X<\infty$ branch and the $-$
root in the $-\infty<X<0$ branch. This situation is quite generic, as we will
see below.

For a particular reduced theory, corresponding to a given branch of the 
original 
higher-derivative theory, we can define $\chi=\log\lm$ and perform the
conformal transformation $\bg_{\mu\nu}=e^\chi g_{\mu\nu}$.
Action \eqref{auxf(R)S} then becomes,
\begin{equation}
\label{canonf(R)}
   S = \frac{1}{2\kappa^2} \int{ d^4x \sqrt{-\bg} \left[ \bR 
           - \tfrac32 \left(\bnabla\chi\right)^2
           - e^{-2\chi} \left( e^\chi X(e^\chi) - f(X(e^\chi)) \right)
           \right] }.
\end{equation}
As promised, we find a theory of Einstein gravity coupled to a scalar field 
with a particular potential dependent on the choice of the original function 
$f(R)$. As in the quadratic case there is one subtlety in defining $\chi$. 
To keep $\chi$ real, we must have $\lambda>0$. When $\lm<0$ we must introduce, 
instead, the field $\chi=\log(-\lm)$, the effect of which is to change the 
sign of the overall normalization of the action as 
compared with the form given in \eqref{canonf(R)} above. Further, at the 
special point $\lm=0$ the action cannot be put in canonical form. In this 
paper we will restrict our attention to the $\lm>0$ case only. 

As a concrete
example, consider $f(R)$ defined in \eqref{eg1}. As pointed out above, in this
case $\lambda>1$ in either branch of $X$. Therefore, $\chi=\log\lm$
is always well defined. We point out, however, that the range of $\chi$ 
is restricted to $0<\chi<\infty$.

Note that the scalar field kinetic energy has the usual sign and, hence, 
$\chi$ is not a ghost. We can now look for the vacua in any given branch of 
the theory. The vacua of the theory described by action \eqref{canonf(R)} 
are defined to be stable solutions of the $\bg_{\mu\nu}$ and $\chi$ equations 
of motion given by
\begin{gather}
\label{canR2geom}
   \bR_{\mu\nu} - \tfrac12\bg_{\mu\nu}\bR = 
         \tfrac32 \left[ \bnabla_\mu\chi\bnabla_\nu\chi
             - \tfrac12 {\bar g}_{\mu\nu} \left(\bnabla\chi\right)^2 \right] - 
         \tfrac32\bg_{\mu\nu}V(\chi) \\
\label{canR2chieom}
   \bnabla^2\chi = \frac{dV}{d\chi} 
\end{gather}
respectively, where the potential is given by
\begin{equation}
\label{Vf(R)}
   V(\chi) = \tfrac13 e^{-2\chi} \left( e^\chi X(e^\chi) - f(X(e^\chi)) 
\right),
\end{equation}
where we recall that the form of $X(e^\chi)$ depends on the branch in 
question. As stated previously, we will only consider vacua satisfying the 
covariant constant condition 
\begin{equation}
\label{covconst}
   \bnabla_\mu\chi = \partial_\mu\chi = 0.
\end{equation}
It follows that $\chi$ is a constant and, from \eqref{canR2chieom}, that it 
must extremize the potential. Furthermore, it follows from \eqref{canR2geom} 
that the vacuum is a space of constant curvature with 
\begin{equation}
   \bR=6V(\chi).
\end{equation}
Since $g_{\mu\nu}=e^{-\chi}\bg_{\mu\nu}$ and $\chi$ is constant, these vacua 
also correspond to spaces of constant curvature with respect to the original 
metric, with $R=e^\chi\bR$. Since the vacuum field $\chi$ must be an extremum 
of the potential, we have the condition
\begin{equation}
\label{minV}
   e^{-2\chi} \left( 2f(X(e^\chi)) - e^\chi X(e^\chi) \right) = 0.
\end{equation}
We started with a completely general function $f$, so this condition may 
generically have multiple solutions, including solutions away from $\chi=0$. 
Further, given a particular solution for $\chi$, the value of the potential 
at this point need not be zero; that is, the vacuum generally has non-zero 
cosmological constant. In this case, the curvature scalars $R$ and $\bR$ will 
be non-vanishing. 

\begin{figure}[ht]
   \centerline{\psfig{figure=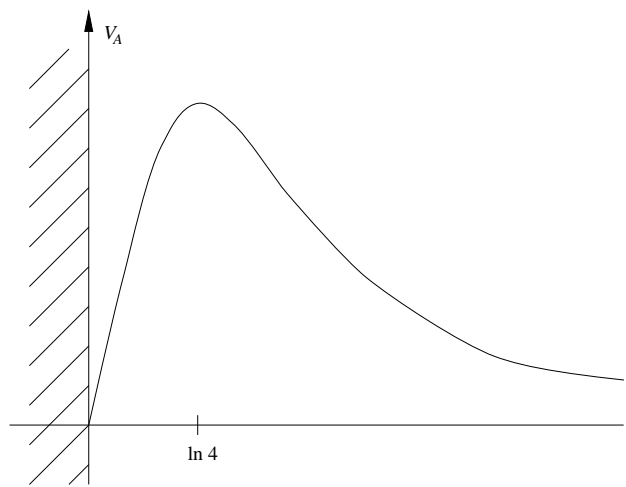,height=2.5in}}
   \caption{$V_A(\chi)$ for $R+\epsilon^{-2}R^3$ gravity}
   \label{fig:R3Vp}
\end{figure}

\begin{figure}[ht]
   \centerline{\psfig{figure=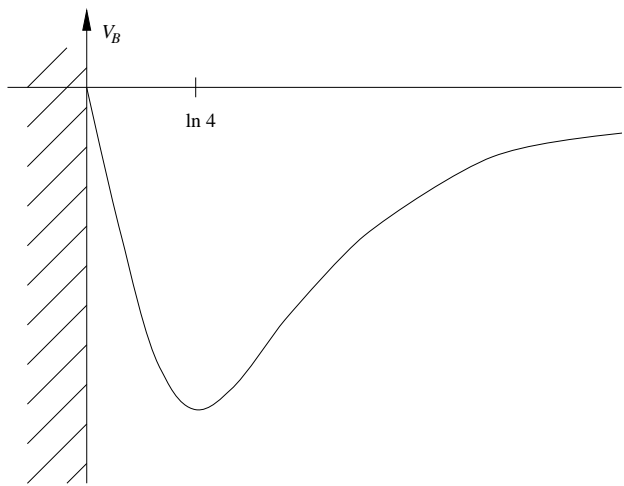,height=2.5in}}
   \caption{$V_B(\chi)$ for $R+\epsilon^{-2}R^3$ gravity}
   \label{fig:R3Vm}
\end{figure}

As a concrete example of all this, we return to the example specified in
\eqref{eg1}. As discussed previously, there are two branches where the second
order theory is defined; branch A where $0<X<\infty$ and branch B where
$-\infty<X<0$. The field $\lambda$, which is given as a function of $X$ in
\eqref{lmeg1}, satisfies $\lambda>1$ everywhere and, hence, $\chi=\log\lambda$
is well defined in both branches, although its range in restricted to
$0<\chi<\infty$. Expression \eqref{lmeg1} was inverted to give $X$ as a 
function of $\lambda$ in \eqref{Xeg1}. This expression is branch dependent, 
being given by $X=\pm(\epsilon/\sqrt{3})\sqrt{\lambda-1}$ with the positive 
root chosen in branch A and the negative root chosen in branch B. It follows 
that the form of the potential energy defined in \eqref{Vf(R)} also depends on 
the branch chosen. It is given by
\begin{equation}
\begin{aligned}
   V_A(\chi) &= +\frac{2}{9\sqrt{3}} \epsilon e^{-2\chi} (e^{\chi}-1)^{3/2} \\
   V_B(\chi) &= -\frac{2}{9\sqrt{3}} \epsilon e^{-2\chi} (e^{\chi}-1)^{3/2}
\end{aligned}
\end{equation}
for branch A and B respectively. The two potentials are plotted in Figures
\ref{fig:R3Vp} and \ref{fig:R3Vm}. We see that each has a stationary point. 
However, since vacua must be  minima of the potential, it follows that only 
branch B has a stable 
vacuum. This is located at $\chi=\ln 4$, which is non-zero. Note that $V_B(\ln 
4)=-\epsilon/24$, so that the vacuum state has a non-vanishing 
negative cosmological constant. It follows that $\bR=-\epsilon/4$ and 
$R=-\epsilon$, which implies that the vacuum state is an anti-de Sitter space 
both in terms of the metric $\bg_{\mu\nu}$ and the metric $g_{\mu\nu}$.
Thus we have given an example of a higher-derivative theory which has a new 
vacuum state away from the flat-space solution of ordinary Einstein gravity. 
It is important to note that none of this structure would have been evident 
if we had made a simple expansion of the original action around flat space 
in terms of $h_{\mu\nu}=g_{\mu\nu}-\eta_{\mu\nu}$. In fact, keeping only the 
first non-trivial, quadratic terms in $h_{\mu\nu}$, we would not even have 
been aware that there was an additional scalar degree of freedom in the 
theory. The lowest order term in $R^3$ is cubic in $h_{\mu\nu}$, and so 
in a quadratic expansion we would only see the usual Einstein term.

The conclusion is that generic higher-derivative 
corrections to pure Einstein gravity introduce completely new 
vacuum states into the theory. Flat space is no longer a unique point in 
field space, so that if, for instance, we wish to investigate the fundamental 
excitations in the theory, we must start by specifying which vacuum state we 
are considering. 

\begin{figure}[ht]
   \centerline{\psfig{figure=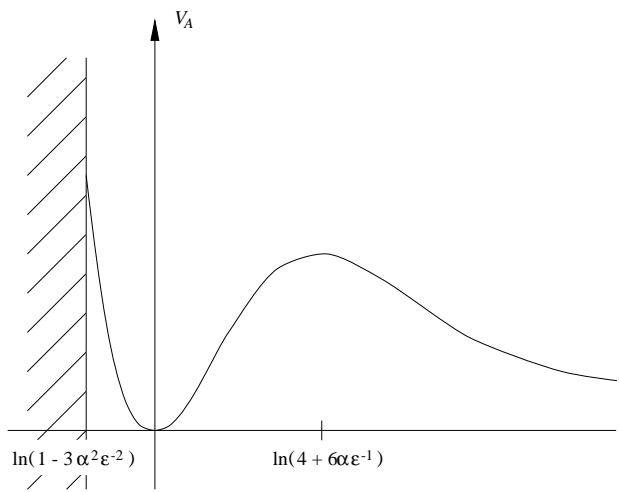,height=2.5in}}
   \caption{$V_A(\chi)$ for $R+3\alpha\epsilon^{-2}R^2+\epsilon^{-2}R^3$ 
       gravity}
   \label{fig:R3genVp}
\end{figure}

\begin{figure}[ht]
   \centerline{\psfig{figure=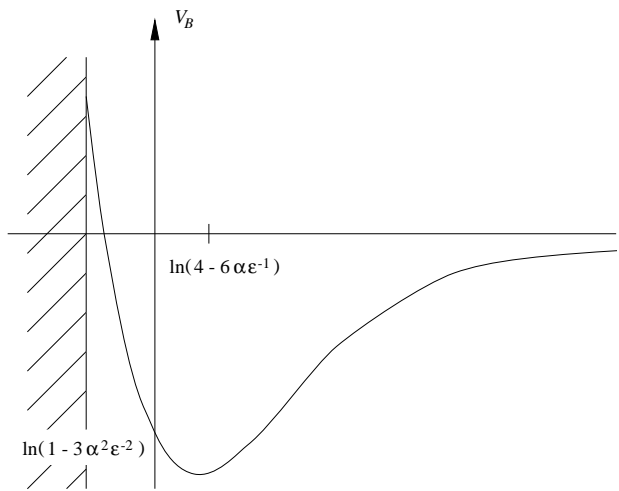,height=2.5in}}
   \caption{$V_B(\chi)$ for $R+3\alpha\epsilon^{-2}R^2+\epsilon^{-2}R^3$ 
       gravity}
   \label{fig:R3genVm}
\end{figure}

At this point, we would like to present another specific example that further
illustrates the preceding discussion. Consider the cubic function
\begin{equation}
   f(R) = R + 3\alpha\epsilon^{-2} R^2+\epsilon^{-2}R^3
\end{equation}
with $\epsilon>\sqrt{3}\alpha>0$. We now have
\begin{equation}
   f''(X)=6\epsilon^{-2}(X+\alpha)
\end{equation}
so that the theory has a single critical point at $X=-\alpha$. It follows that
the second-order theory is defined on two branches; branch A where
$-\alpha<X<\infty$ and branch B where $-\infty<X<-\alpha$. On either branch we
can define a new field $\lambda=f'(X)$, which in this case is given by
\begin{equation}
\label{lmeg2}
    \lambda = 1+6\alpha\epsilon^{-2}X+3\epsilon^{-2}X^2.
\end{equation}
Note that, since $\epsilon>\sqrt{3}\alpha$, $\lambda$ is always a positive
real number in the range $1-3\alpha^2\epsilon^{-2}<\lambda<\infty$ for both
branches. Expression \eqref{lmeg2} can be inverted to give
\begin{equation}
\label{Xeg2}
    X_{\pm} = -\alpha \pm \sqrt{\alpha^2+\tfrac13\epsilon^2(\lambda-1)}.
\end{equation}
Clearly the $X_+$ solution is correct on branch A whereas the $X_-$ solution
is to be used in branch B. Since $\lambda$ is always positive, the definition
of $\chi=\ln\lm$ is valid in both branches. Note that $\chi$ is then 
restricted to lie in the range
$\ln(1-3\alpha^2\epsilon^{-2})<\chi<\infty$. Since the expression for $X$ in
terms of $\lambda$, and hence $\chi$, given in \eqref{Xeg2} is
different in each branch, it follows from \eqref{Vf(R)} that the
potential energy function also is different in each branch. We have
\begin{equation}
\begin{aligned}
   V_A(\chi) &= \tfrac13 \epsilon^{-2} e^{-2\chi} 
                 \left(X_+(\chi)\right)^2\left(3\alpha+2X_+(\chi)\right), \\
   V_B(\chi) &= \tfrac13 \epsilon^{-2} e^{-2\chi} 
                 \left(X_-(\chi)\right)^2\left(3\alpha+2X_-(\chi)\right), 
\end{aligned}
\end{equation}
where
\begin{equation}
   X_\pm(\chi) = -\alpha \pm \sqrt{\alpha^2+\tfrac13\epsilon^2(e^\chi-1)}
\end{equation}
and the $V_A$ expression is valid in branch A and the $V_B$ expression valid 
in branch B. The two potentials are plotted in Figures \ref{fig:R3genVp} and 
\ref{fig:R3genVm}. Unlike the previous example, in this case each branch 
contains a single, stable minimum. In branch A, this minimum is located at 
$\chi=0$ and has vanishing cosmological constant $V(0)=0$. It follows that, 
for this vacuum, $\overline{R}=R=0$ and spacetime is flat with respect to 
both the $\bg_{\mu\nu}$ and $g_{\mu\nu}$ metrics. In branch B, the minimum is 
located at $\chi=\ln(4-6\alpha\epsilon^{-1})$ with a non-vanishing negative 
cosmological constant given by $V(\chi)=-\epsilon^2/6(4\epsilon-6\alpha)$. 
For this vacuum $\bR=-\epsilon^2/(4\epsilon-6\alpha)$
and, hence, $R=-\epsilon$. It follows that the spacetime is an anti de-Sitter
space with respect to both metrics $\bg_{\mu\nu}$ and $g_{\mu\nu}$. 
Thus we have found an example which, aside from a conventional minimum at 
$R=0$
with zero cosmological constant, has an additional minimum with negative
cosmological constant. Furthermore, there is an unstable maximum with positive
cosmological constant. The mass of the scalar field is different at each of 
the
different minima. Here a quadratic expansion of the action around flat-space
would have identified a new scalar degree of freedom, but would never have
revealed the presence of a second stable vacuum state.

It will not have escaped the readers notice that finding the vacua in the 
second-order formalism requires a careful discussion of the branches in the 
$X$
variable. The branching structure was not too difficult in the preceding
examples, but as we will show below, it can be, and generally is, extremely
complicated. However, the following remarks will allow us to define a simpler
procedure for determining the vacua. Recall that, for a covariantly constant
scalar $\chi$ satisfying $dV/d\chi=0$, the corresponding spacetime structure
in terms of the metric $g_{\mu\nu}$ is that of a space of constant curvature 
with $R=6e^\chi V(\chi)$. Provided we are not at a degenerate point, spaces of
constant curvature $R$ and vacuum solutions are in one-to-one correspondence,
since the relation $R=6e^\chi V(\chi)$ not only means that constant $\chi$
implies constant $R$, but also the reverse. Note that, as a corollary, any 
vacuum solution with zero cosmological constant, namely $V(\chi)=0$, must 
correspond to flat space in the original higher-derivative theory. An 
important consequence of this result is that, since flat space is a single 
point in the field space of the original theory, there can be only one vacuum 
state with zero cosmological constant in the second-order theory. We conclude 
that all vacuum solutions of the above type can be found as constant curvature
solutions of the equations of motion for the higher-derivative $g_{\mu\nu}$ 
theory. All such vacua can be found directly, without reference to any branch 
structure. Having found a vacuum of interest, one can then proceed in reverse, 
introducing the second-order theory in the appropriate branch containing that 
vacuum. This is often an easier procedure. The $g_{\mu\nu}$ equations derived 
from the higher-derivative action \eqref{f(R)S} are 
\begin{equation}
   R_{\mu\nu}f'(R) - \tfrac12g_{\mu\nu}f(R)
       - \left( \nabla_\mu\nabla_\nu - g_{\mu\nu}\nabla^2 \right) f'(R) = 0.
\end{equation}
If we look for solutions of constant curvature, then
$R_{\mu\nu}=\tfrac14g_{\mu\nu}R$ where $R$ is a constant. It follows that the 
derivative 
terms in the equations of motion drop out and we are left with the simple 
condition, first derived by Barrow and Ottewill~\cite{JPA-16-2757},
\begin{equation}
\label{constR}
   Rf'(R) - 2f(R) = 0.
\end{equation}
We see that this is exactly the condition 
we obtained for stationary points of the potential \eqref{minV} aside from a 
factor of $e^{-2\chi}$. In the latter 
case, however, the expression was taken to be a function of the scalar field 
$\chi$ expressed in terms of $X$, and as such only valid in a branch-by-branch 
sense, whereas here the expression is valid for all curvature $R$.

This now provides us with a procedure for finding all the vacua of the 
theory as well as the nature of the excitations around a given vacuum. 
We start by looking for constant curvature solutions of the original 
higher-derivative equations of motion, solving the equation \eqref{constR}, 
which is valid globally. Having identified these points we can then, 
locally around each solution, make a transformation to the reduced theory, 
choosing, if necessary, the appropriate branch of $X(e^\chi)$. In the 
transformed 
frame, the new scalar degree of freedom is made explicit. We can then address 
questions of local stability and the mass of the scalar field. The only 
breakdown of this procedure occurs when the minimum is at a degenerate point 
of the transformation; that is, a point where $f''(R)=0$. In this case, any 
discussion of the particle spectrum and local stability must be in terms of 
the 
original theory. In general, we shall not consider such degenerate points. 

\begin{figure}[ht]
   \centerline{\psfig{figure=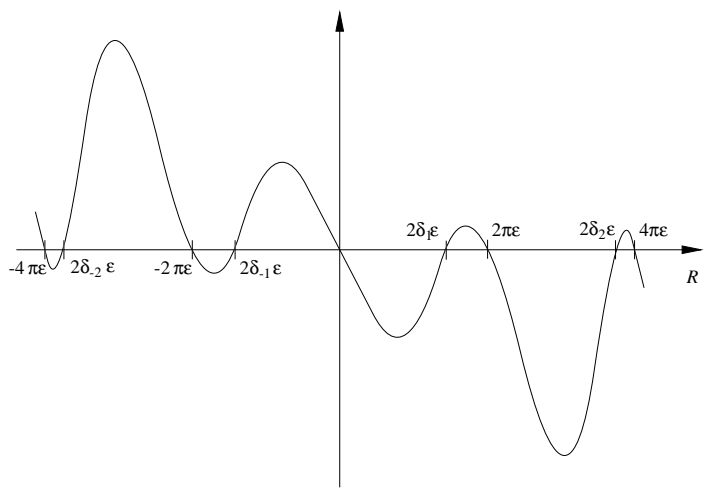,height=2.5in}}
   \caption{Condition for a constant curvature solution of 
       $\tfrac12\left[R+\epsilon\sin (R/\epsilon)\right]$ gravity}
   \label{fig:constRex}
\end{figure}

As an example of this procedure, consider the function
\begin{equation}
   f(R) = \tfrac12 \left[ R + \epsilon \sin\left(R/\epsilon\right) \right].
\end{equation}
Where $\epsilon$ is a constant. The condition for a degenerate point is then
\begin{equation}
   f''(R) = - \tfrac12 \epsilon^{-1} \sin\left(R/\epsilon\right) = 0.
\end{equation}
This equation has solutions $R=n\pi\epsilon$ where $n$ is any integer. The 
theory thus has an infinite number of branches given by 
$n\pi\epsilon<R<(n+1)\pi\epsilon$. Rather than treating each branch 
separately, 
we follow our procedure for finding the vacuum states of the theory by 
solving the condition \eqref{constR} for the constant curvature solutions of 
the original theory. The condition reads
\begin{equation}
\label{constRex}
   \sin\left(R/2\epsilon\right) \left[
        \left(R/2\epsilon\right)\sin\left(R/2\epsilon\right)
        - \cos\left(R/2\epsilon\right) \right] = 0.
\end{equation}
The left-hand side of this expression is plotted in figure \ref{fig:constRex}. 
We have the solutions
\begin{equation}
   R = 2n\pi\epsilon, \; 2\delta_n\epsilon
\end{equation}
where $n$ is an integer and $\delta_n$ are the solutions of
\begin{equation}
\label{deltaeq}
   \delta\sin\delta - \cos\delta = 0.
\end{equation}
Explicitly, the first few solutions of \eqref{deltaeq} near $\delta=0$ are 
\begin{equation}
   \dots, \; \delta_{-2} \approx -6.12, \; \delta_{-1} \approx -2.80, \; 
   \delta_1 \approx 2.80, \; \delta_2 \approx 6.12, \; \dots
\end{equation}
Furthermore 
$(\left|n\right|-\tfrac12)\pi<\left|\delta_n\right|<\left|n\right|\pi$ and as 
$\left|\delta\right|$ becomes large the solutions approach $\delta_n=n\pi$.  

The first observation to make is that the constant curvature solutions at 
$R=2n\pi\epsilon$, including the point $R=0$, all correspond to degenerate 
points. As such, a second-order form of the theory does not exist at these 
points. Thus, although these are constant curvature solutions, we 
cannot identify them as vacuum solutions of a canonical second-order theory. 
However the solutions at $R=2\delta_n\epsilon$ are not at degenerate points. 
From our previous discussion this implies that they must correspond to vacuum 
solutions of the canonical second-order theory. Nonetheless, we find that each 
vacuum solution lies in a different branch of the theory. 

\begin{figure}[ht]
   \centerline{\psfig{figure=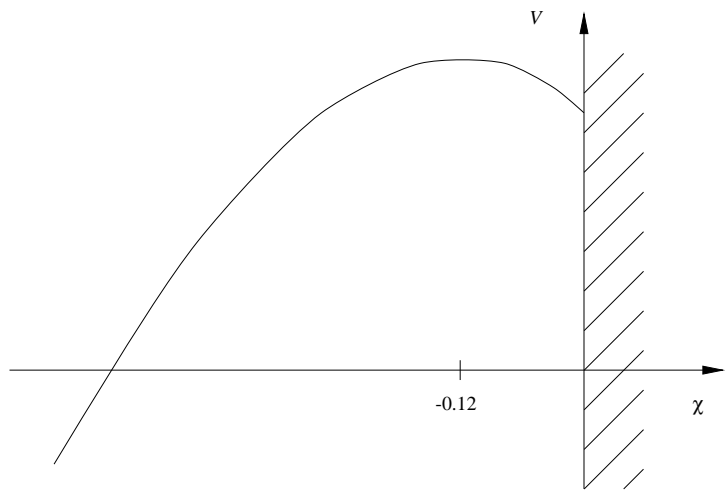,height=2.5in}}
   \caption{$V(\chi)$ for the branch $\pi\epsilon<R<2\pi\epsilon$}
   \label{fig:sinRVp}
\end{figure}

Let us consider two particular solutions,
$R=2\delta_1\epsilon\approx 5.60\epsilon$ and
$R=2\delta_{-1}\epsilon=-2\delta_1\epsilon\approx -5.60\epsilon$, in the 
second-order formulation. For the first solution 
we must take the branch $\pi\epsilon<R<2\pi\epsilon$. Following 
our usual procedure for writing the theory in a second-order form, we first 
introduce the auxiliary field $X$ and then define
\begin{equation}
   \lm = f'(X) = \tfrac12 \left[ 1 + \cos\left(X/\epsilon\right) \right].
\end{equation}
However, as usual, we are required to invert this expression to give $X$ as a 
function of $\lm$. It follows that
\begin{equation}
   X = \epsilon \cos^{-1}\left(2\lm-1\right) 
     \qquad \text{with $\pi\epsilon < X < 2\pi\epsilon$}.
\end{equation}
We note that $\lm$ is only defined in the range $0<\lm<1$. To put the action 
in canonical second-order form we define $\chi=\ln\lm$. The potential for 
$\chi$ \eqref{Vf(R)}, is then given by
\begin{equation}
\label{V1ex3}
   V(\chi) = \tfrac13 e^{-2\chi} \left[
        X(\chi) \cos\left(X(\chi)/\epsilon\right) 
        - \epsilon\sin\left(X(\chi)/\epsilon\right) \right]
\end{equation}
where,
\begin{equation}
   X(\chi) = \epsilon \cos^{-1}\left(2e^\chi-1\right)
     \qquad \text{with $\pi\epsilon < X < 2\pi\epsilon$}.
\end{equation}
We now have that $\chi$ is restricted to the range $-\infty<\chi<0$. The 
potential is plotted in figure \ref{fig:sinRVp}. We find that there is a 
single maximum of the potential at $\chi\approx-0.12$. The value of the 
potential at this point is $V(\chi)\approx1.05\epsilon$, so that 
$\bR=6V(\chi)\approx6.31\epsilon$ and 
$R=6e^\chi V(\chi)\approx5.60\epsilon$. Thus we see that the maximum 
corresponds to the flat space solution $R=2\delta_1\epsilon$ as expected, 
and we can conclude that this vacuum is unstable.

\begin{figure}[ht]
   \centerline{\psfig{figure=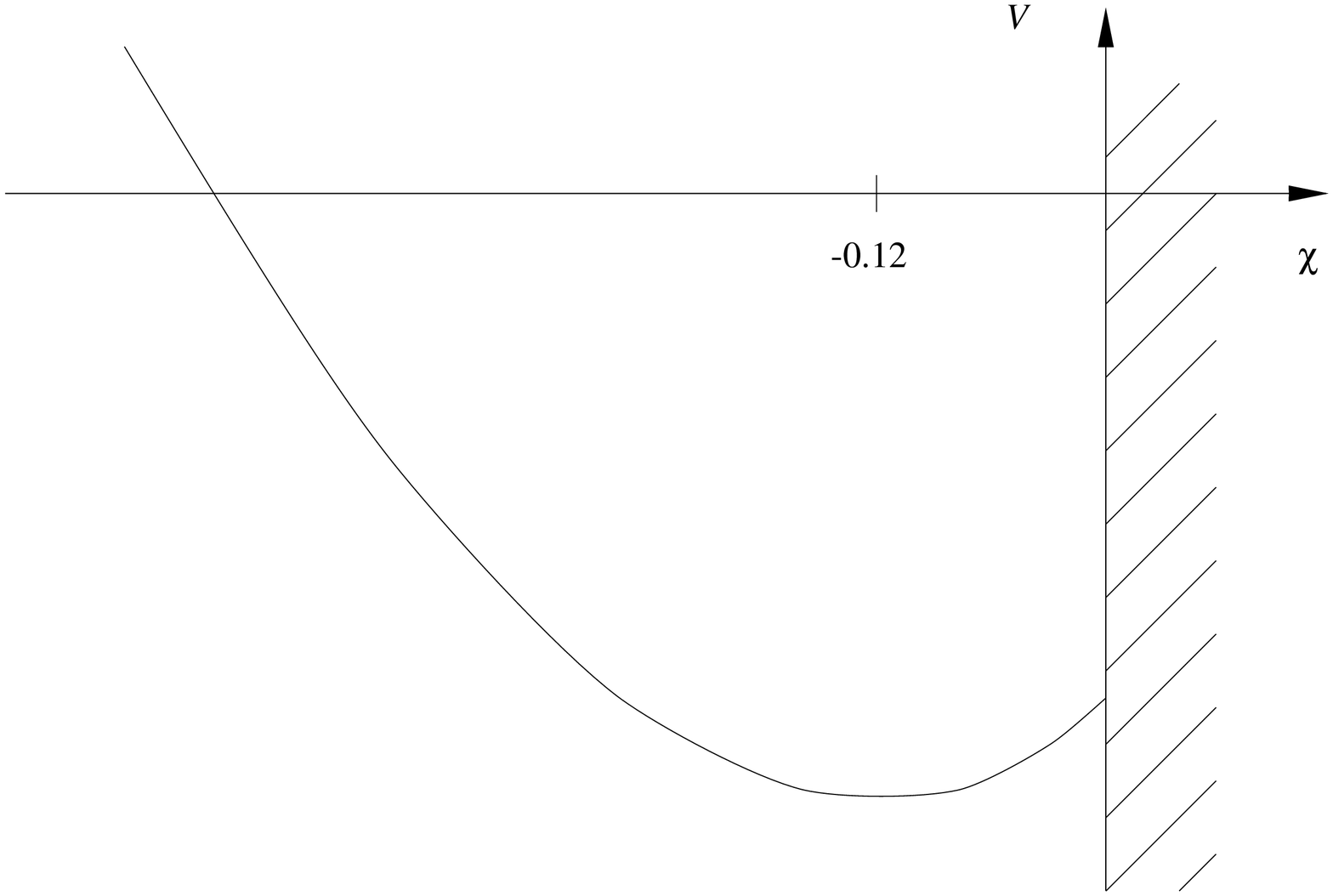,height=2.5in}}
   \caption{$V(\chi)$ for the branch $-2\pi\epsilon<R<-\pi\epsilon$}
   \label{fig:sinRVm}
\end{figure}

For the solution at $R=2\delta_{-1}\epsilon$, we must take the branch 
$-2\pi\epsilon<R<-\pi\epsilon$. The expression for the potential is the 
same as the expression \eqref{V1ex3} above, except that we must now define 
\begin{equation}
   X(\chi) = \epsilon \cos^{-1}\left(2e^\chi-1\right)
     \qquad \text{with $-2\pi\epsilon < X < -\pi\epsilon$}.
\end{equation}
The resulting potential is plotted in figure \ref{fig:sinRVm}. Again $\chi$ 
is restricted to lie in the range $-\infty<\chi<0$. We now find a 
stable minimum at $\chi\approx-0.12$. This gives 
$V(\chi)\approx-1.05\epsilon$, $\bR=6V(\chi)\approx-6.31\epsilon$ and 
$R=6e^\chi V(\chi)\approx-5.60\epsilon$, showing that the minimum of the 
potential does indeed correspond to the constant curvature solution
$R=2\delta_{-1}\epsilon$. We can concluded that this is a stable vacuum 
state. In fact, further analysis shows that all the vacua 
$R=2\delta_n\epsilon$ with $\delta_n>0$ are unstable maxima, while those 
with $\delta_n<0$ are stable minima. This final example demonstrates that 
the vacuum structure of higher-derivative theories can be complicated. We 
find an infinite number of stable and unstable vacua, each in a different 
branch of the theory.  

In summary, we have shown that theories which are general functions of the 
scalar curvature $R$, can be reduced to a canonical second-order form 
describing Einstein gravity coupled to a scalar field with a potential which 
may have a vacuum structure of arbitrary complexity. Further these vacua all 
correspond to constant curvature solutions of the original theory. The 
advantage of discussing the theory in the second-order formalism is that it is 
in a canonical form where we can easily identify the excitations around the 
vacuum. The disadvantage is that the problem of inverting $f'(X)$ means that 
we 
are forced to break the theory into regions separated by degenerate points. 
For this reason it is often easier to identify vacua as constant curvature 
solutions of the higher-derivative theory and then make a local transformation 
in the region of each solution to investigate the excitations around that 
vacuum. All of the non-trivial vacuum structure arises from non-linear terms 
in the field equations and so is missed in a quadratic expansion of the 
higher-derivative theory in terms of the fluctuation of the metric around 
flat space. Further, any analysis of the particle content of the theory, in 
this case the presence of a new scalar degree of freedom and identification 
of its mass, can only be made once the vacuum is identified.

\section{Actions Given by General Functions of the Ricci Tensor}

In this and the following section we shall extend our discussion to include 
actions which are general functions of first the Ricci tensor and then the 
Riemann tensor. We shall use the same techniques as in the previous section,
namely introducing auxiliary fields, showing that vacuum states
correspond to constant curvature solutions of the higher derivative
theory, and then investigating  
the excitations around a given vacuum. The reduction to a second-order
form was first discussed by Magnano 
\textit{et al.}~\cite{GRG-19-465,CQG-5-L95,CQG-7-557}
and Jakubiec and Kijowski~\cite{PRD-37-1406}. Here, we shall
use a slightly different analysis, again pointing out the need for
identifying different branches of the theory and also giving a
canonical separation of the new variables about any given vacuum. We
shall only give the general formulation without specific
examples. Again we will find a rich structure of vacua which would be
completely missed in a linearized analysis. This will re-emphasize the
need to identify the vacuum in question before investigating the
masses and couplings of the elementary excitations of the theory. 

Starting with an action $f(R_{\mu\nu})$ we can introduce an auxiliary field 
and put the theory in a second-order form 
\begin{align}
   S &= \frac{1}{2\kappa^2} \int{ d^4x \sqrt{-g} f(R_{\mu\nu}) } \notag \\
     &= \frac{1}{2\kappa^2} \int{ d^4x \sqrt{-g} \left[
            \frac{df(X_{\rh\sg})}{dX_{\mu\nu}} 
            \left(R_{\mu\nu} - X_{\mu\nu}\right) + f(X_{\mu\nu}) \right]
            } \notag \\
     &= \frac{1}{2\kappa^2} \int{ d^4x \sqrt{-g} \left[ 
            \pi^{\mu\nu} \left(R_{\mu\nu} - X_{\mu\nu}(\pi_{\rh\sg})\right) 
            + f(X_{\mu\nu}(\pi_{\rh\sg})) \right] }.
\end{align}
Again we introduce the auxiliary field in two steps, first writing the action 
in terms of the field $X_{\mu\nu}$ which gets set to $R_{\mu\nu}$ on
solving its equation of motion. We then define 
\begin{equation}
\label{defpi}
   \pi^{\mu\nu} = \frac{df(X_{\rh\sg})}{dX_{\mu\nu}}, 
\end{equation}
and invert the expression to give $X_{\mu\nu}$ as a function of 
$\pi_{\mu\nu}$. Note again that the introduction of both $X_{\mu\nu}$
and $\pi_{\mu\nu}$ requires the non-degeneracy condition
\begin{equation}
   \det\frac{d^2f(X_{\kappa\lm})}{dX_{\mu\nu}dX_{\rh\sg}} \neq 0
\end{equation}
if the auxiliary field is to be properly eliminated to return to the
original action. Furthermore, in inverting \eqref{defpi}, it may be
necessary to divide the theory into branches corresponding to different
possible roots for $X_{\mu\nu}$. Thus for the variable 
$\pi_{\mu\nu}$ it may be necessary to introduce a collection of
auxiliary variable theories each valid for a different branch of 
$R_{\mu\nu}$.

We can ask how many degrees of freedom are there in the auxiliary field? We 
have argued that we expect six. If we consider the $X_{\mu\nu}$ formulation 
for now, we see first that we can again derive a spin-two divergence 
condition. Since the auxiliary field equation of motion gives 
$X_{\mu\nu}=R_{\mu\nu}$, we clearly can show, using the Bianchi identity 
$\nabla^\mu G_{\mu\nu}=0$, that 
$\nabla^\mu\left(X_{\mu\nu}-\tfrac12g_{\mu\nu}X\right)=0$.
Constraining the components of $X_{\mu\nu}$, these four conditions
imply that we do indeed have six new propagating degrees of freedom. We would
like to be able to separate these into scalar and spin-two degrees of
freedom as was done in the quadratic case in \cite{PRD-53-5583}. For a general 
function $f$, this is not possible since we are not able to obtain the 
necessary trace condition from the $g_{\mu\nu}$ equation of
motion. However, as we will show below, having identified a vacuum state, we
can always make the separation locally in an expansion around the vacuum 
solution. 

What then are the vacua of the $f(R_{\mu\nu})$ theory? Since we are unable to
separate the spin-two and scalar degrees of freedom in the auxiliary
field, we cannot put the theory in a canonical form and look to minimize
the potential as we did for the $f(R)$ theory. However, we recall that
the vacuum states in question are none other than states where the
auxiliary field is covariantly constant. This means, since the $X_{\mu\nu}$
equation of motion gives $X_{\mu\nu}=R_{\mu\nu}$ that they are states of 
constant tensor curvature. We can also impose the condition that the states
should have no preferred direction, so that $X_{\mu\nu}$ is proportional 
to the metric $g_{\mu\nu}$. Thus vacuum states are states of constant scalar 
curvature $R_{\mu\nu}=\tfrac14R g_{\mu\nu}$ with $R$ constant. 
Consequently, one approach to finding vacuum solutions is simply to 
look for constant curvature solutions of the original theory. 

The equations of motion of the original higher derivative theory are given by
\begin{equation}
   \nabla^2f'_{\mu\nu} + g_{\mu\nu}\nabla^\rh\nabla^\sg f'_{\rh\sg}
   - \nabla_\rh\nabla_\mu{f'_\nu}^\rh - \nabla_\rh\nabla_\nu{f'_\mu}^\rh
   + R_{\rh\mu}{f'_\nu}^\rh + R_{\rh\nu}{f'_\mu}^\rh 
   - g_{\mu\nu} f = 0.
\end{equation}
To incorporate the presence of the metric in $f(R_{\mu\nu})$ in deriving the 
equations of motion, we consider $f$ as a function of the mixed index object
${R_\mu}^\nu$, with all contractions made between raised and lowered indices 
so 
that the metric does not enter explicitly. Then in the above expression
${f'_\mu}^\nu=df/d{R_\nu}^\mu$, so that, for instance, 
$f'_{\mu\nu}={f'_\mu}^\lm g_{\lm\nu}$. Looking for constant curvature 
solutions of the form $R_{\mu\nu}=\tfrac14g_{\mu\nu}R$, we obtain the 
condition
\begin{equation}
\label{vaccond}
   R g'(R) - 2g(R) = 0,
\end{equation}
where $g(R)=f(\tfrac14R{\delta_\mu}^\nu)$ and
$g'(R)=dg(R)/dR$, and we have used the fact that ${f'_\mu}^\nu$
evaluated at the constant curvature solution must be proportional to
${\delta_\mu}^\nu$. Note that this condition has exactly the same form as 
the condition \eqref{constR} we obtained for actions which were general 
functions of the scalar curvature.

Given a particular vacuum solution satisfying the condition \eqref{vaccond}, 
we would like to investigate the excitations around 
the vacuum state. The natural way to do this is to consider an expansion of 
the action around a given vacuum $R=R_0$. Expanding $f(R_{\mu\nu})$ and 
keeping terms to quadratic order in the curvature only, we have 
\begin{align}
   S &= \frac{1}{2\kappa^2} \int{ d^4x \sqrt{-g} \left[
          a_0 + a_1 \left(R-R_0\right) + \tfrac12 a_{2,1} \left(R-R_0\right)^2
          \right. } \notag \\
          & { \left. \hspace*{1.5in} + \tfrac12 a_{2,2} \left(R_{\mu\nu} 
            - \tfrac14g_{\mu\nu}R_0\right)
          \left(R^{\mu\nu}-\tfrac14 g^{\mu\nu}R_0\right) \right] } \notag \\
     &= \frac{\beta}{2\kappa^2} \int{ d^4x \sqrt{-g} \left[
          - \tfrac12 R_0 + R + \frac{1}{6{m_0}^2}R^2
          - \frac{1}{{m_2}^2}\left(R_{\mu\nu}R^{\mu\nu}
              -\tfrac13R^2\right) \right] }, 
\label{expandf}
\end{align}
Here we have evaluated $f$ and its derivative at the vacuum 
$R_{\mu\nu}=\tfrac14g_{\mu\nu}R_0$ and defined
\begin{equation}
\begin{aligned}
   \left. f \right|_{R_0} &= a_0,  \\
   \left. \frac{df}{dR_{\mu\nu}} \right|_{R_0} &= a_1 g^{\mu\nu},  \\
   \left. \frac{d^2f}{dR_{\mu\nu}dR_{\rh\sg}} \right|_{R_0} &= 
        a_{2,1} g^{\mu\nu}g^{\rh\sg} + \tfrac12 a_{2,2} 
        \left( g^{\mu\rh}g^{\nu\sg} + g^{\nu\rh}g^{\mu\sg} \right), 
\end{aligned}
\end{equation}
using the general symmetric decomposition of the second two expressions. The 
variables $\beta$, $m_0$ and $m_2$ are then given by 
\begin{equation}
\begin{aligned}
   \beta &= a_1 - \left(a_{2,1} + \tfrac14a_{2,2}\right)R_0 , \\
   {m_0}^2 &= \frac{\beta}{3a_{2,1}+a_{2,2}}, \\
   {m_2}^2 &= -\frac{2\beta}{a_{2,2}}, 
\end{aligned}
\end{equation}
where in writing the final line of \eqref{expandf} we have used the
constant curvature condition $a_1R_0=2a_0$.
We see that we have succeeded in putting the action in a quadratic
form, though with the addition of a cosmological constant. 

The final line of \eqref{expandf} is identical to the general quadratic form 
we discussed in our previous paper \cite{PRD-53-5583}, except for the addition 
of 
a cosmological constant term and a  renormalization of the gravitational 
coupling constant by a factor $\beta$. We can thus introduce auxiliary 
fields $\chi$ and $\tpi_{\mu\nu}$ exactly as we did in \cite{PRD-53-5583} to 
give
\begin{multline}
\label{transfRicci}
   S = \frac{\beta}{2\kappa^2} \int{ d^4x \sqrt{-\tg}} \left[ \tR 
          - \tfrac32\left(\tnabla\chi\right)^2
          - \tfrac32{m_0}^2\left(1-e^{-\chi}\right)^2
          - \tfrac12R_0 e^{-2\chi} 
          \right. \\ \left.
          - \tG_{\mu\nu}\tpi^{\mu\nu}
          + \tfrac14{m_2}^2\left(\tpi_{\mu\nu}\tpi^{\mu\nu}-\tpi^2\right)
          \right]. 
\end{multline}
The only effect of the cosmological term is to modify the potential for the
scalar field, adding a term $\tfrac12R_0e^{-2\chi}$. We find that the 
extremum of the potential is now at $3{m_0}^2\left(e^\chi-1\right)=R_0$, 
which, relating the auxiliary field $\chi$ back to the original curvature, 
gives $R=R_0$, as required for a consistent expansion. Following exactly the 
analysis of our previous paper, the auxiliary field $\tpi_{\mu\nu}$ satisfies 
generalized divergence and trace conditions, so it does indeed describe the 
degrees of freedom of a spin-two field. By making a final field redefinition, 
we can write the action in canonical form with explicit kinetic energy terms 
for the spin-two field. Again following the discussion in our previous paper, 
the spin-two field will have the correct Pauli-Fierz limit, but will 
unfortunately be ghost-like.

Thus we have shown that locally, around any vacuum of the
higher-derivative theory (that is a solution of constant curvature), we
can expand the theory to identify a new scalar and a new spin-two
degree of freedom (provided we are not at a degenerate point). The spin-two 
field satisfies divergence and trace conditions as before but importantly, we 
see that it remains ghost-like. Thus generalizing to $f(R_{\mu\nu})$ actions 
fails to remove the problem of the ghost spin-two degree of freedom. The
masses of the degrees of freedom are fixed by the form of the function
$f$ around the constant curvature solution. 

It is worth noting that we could equally well have done this analysis in the 
second-order form, looking for solutions with constant $X_{\mu\nu}$ 
proportional to $g_{\mu\nu}$. Then expanding in small $X_{\mu\nu}$ about 
such solutions, gives a linear coupling between $X_{\mu\nu}$ and $R_{\mu\nu}$ 
and quadratic `mass' terms for $X_{\mu\nu}$. From this form we could then 
extract the scalar and spin-two parts of $X_{\mu\nu}$. In this sense, the 
auxiliary field $X_{\mu\nu}$ always carries six degrees of freedom, which 
about any given vacuum can be decomposed into a scalar field and a ghost 
spin-two field, with the form of the decomposition changing as we go from 
vacuum to vacuum.

\section{Actions Given by General Functions of the Riemann Tensor}

Our last generalization is to consider actions which are general functions 
of the Riemann tensor $R_{\lm\mu\nu\rh}$. As mentioned earlier, the suggestion 
is that such theories have an additional six degrees of freedom,
which we know in the quadratic case can be decomposed into scalar and
spin-two fields. 

We start the discussion by demonstrating that, as in all previous cases,
we can introduce an auxiliary field to write the action in a second
order form. 
\begin{align}
   S &= \frac{1}{2\kappa^2} \int{ d^4x \sqrt{-g} f(R_{\lm\mu\nu\rh}) 
            } \notag \\
     &= \frac{1}{2\kappa^2} \int{ d^4x \sqrt{-g} \left[
            \frac{df(X_{\alpha\beta\gamma\delta})}{dX_{\lm\mu\nu\rh}} 
            \left(R_{\lm\mu\nu\rh} - X_{\lm\mu\nu\rh}\right) 
            + f(X_{\lm\mu\nu\rh}) \right] } \notag \\
     &= \frac{1}{2\kappa^2} \int{ d^4x \sqrt{-g} \left[ 
            \rho^{\lm\mu\nu\rh} \left(R_{\lm\mu\nu\rh} 
                 - X_{\lm\mu\nu\rh}(\rho_{\alpha\beta\gamma\delta})\right) 
            + f(X_{\lm\mu\nu\rh}(\rho_{\alpha\beta\gamma\delta})) \right] }.
\end{align}
Again we introduce the auxiliary field in two stages; first introducing
$X_{\lm\mu\nu\rh}$, which gets set equal to the Riemann tensor on
solving its equation of motion, and then defining
\begin{equation}
   \rho^{\lm\mu\nu\rh} = 
         \frac{df(X_{\alpha\beta\gamma\delta})}{dX_{\lm\mu\nu\rh}} 
\end{equation}
In both cases we require the non-degeneracy condition
\begin{equation}
   \det \frac{d^2f(X_{\alpha\beta\gamma\delta})}
          {dX_{\eta\kappa\lm\mu}dX_{\nu\rh\sg\tau}} = 0,
\end{equation}
in order to be able to eliminate the auxiliary field and return to the 
original 
action.  When using the variable $\rho_{\lm\mu\nu\rh}$ we may be required 
to break the theory into branches, introducing a collection of second-order 
theories, each taking a different branch when inverting to find 
$X_{\lm\mu\nu\sg}$ in terms of $\rho_{\lm\mu\nu\rh}$. 

Turning to the vacuum states, we again find that states with constant 
$X_{\lm\mu\nu\sg}$ have constant Riemann curvature, since $X_{\lm\mu\nu\sg}$ 
is 
set equal to $R_{\lm\mu\nu\sg}$ by its equation of motion. Furthermore, 
imposing 
the condition that there is no preferred direction in spacetime we require 
that
$X_{\lm\mu\nu\sg}$ is proportional to $g_{\mu\nu}$. Therefore, given the 
symmetries of $X_{\lm\mu\nu\sg}$, we have 
\begin{equation}
\label{constRsol}
   R_{\lm\mu\nu\sg} = R^0_{\lm\mu\nu\sg} = \tfrac{1}{12} R_0 \left( 
               g_{\lm\nu}g_{\mu\sg} - g_{\lm\sg}g_{\mu\nu} \right), 
\end{equation}
and, hence, we are considering solutions of constant Ricci scalar 
curvature.

As before the easiest way to obtain such solutions is from the equations of 
motion of the original higher-derivative action. Again, to circumvent the 
problem, when deriving the equation of motion, of the metric explicitly 
entering the function $f$, we consider $f$ as a function of 
$R_{\mu\;\:\nu}^{\;\;\rh\;\:\sg}$ with all contractions made between 
raised and lowered indices. We then derive the equations of motion
\begin{equation}
   \nabla^\rh\nabla^\sg f'_{\mu\rh\nu\sg}
   + \nabla^\rh\nabla^\sg f'_{\nu\rh\mu\sg}
   + \tfrac12 {R_\mu}^{\rh\sg\tau} f'_{\nu\rh\sg\tau}
   + \tfrac12 {R_\nu}^{\rh\sg\tau} f'_{\mu\rh\sg\tau}
   - \tfrac12 g_{\mu\nu} f = 0, 
\end{equation}
where 
${f'}_{\mu\;\:\nu}^{\;\;\rh\;\:\sg}=df/dR_{\rh\;\:\sg}^{\;\;\mu\;\:\nu}$. 
Restricting to constant curvature solutions of the form \eqref{constRsol}, we 
get the familiar condition 
\begin{equation}
   R_0 g'(R_0) - 2g(R_0) = 0,
\end{equation}
where now $g(R_0)=f(R^0_{\lm\mu\nu\rh})$ and $g'(R_0)=dg(R_0)/dR_0$, and we 
have used the fact that, by symmetry, 
${f'}_{\mu\;\:\nu}^{\;\;\rh\;\:\sg}(R^0_{\lm\mu\nu\rh})$ is proportional to 
${\delta_\mu}^\rh{\delta_\nu}^\sg-g_{\mu\nu}g^{\rh\sg}$. 

To investigate the excitations around a given vacuum we expanding the action 
about the constant curvature solution. We have, keeping terms up to quadratic 
order in the curvature only,
\begin{align}
   S &= \frac{1}{2\kappa^2} \int{ d^4x \sqrt{-g} \left[
           a_0 + a_1 \left(R-R_0\right) 
        \right. } \notag \\ & \qquad\qquad \left.
           + \tfrac12 a_{2,1} \left(R-R_0\right)^2
           + \tfrac12 a_{2,2} \left(R_{\mu\nu}-\tfrac14g_{\mu\nu}R_0\right)
           \left(R^{\mu\nu}-\tfrac14 g^{\mu\nu}R_0\right) 
        \right. \notag \\ & \qquad\qquad \left.
           + \tfrac12 a_{2,3} \left( R_{\lm\mu\nu\rh} - \tfrac{1}{12}R_0 
               \left[g_{\lm\nu}g_{\mu\rh}-g_{\lm\rh}g_{\mu\nu}\right] \right)
             \left( R^{\lm\mu\nu\rh} - \tfrac{1}{12}R_0 
               \left[g^{\lm\nu}g^{\mu\rh}-g^{\lm\rh}g^{\mu\nu}\right] \right)
           \right] \notag \\
     &= \frac{\beta}{2\kappa^2} \int{ d^4x \sqrt{-g} \left[
           - \tfrac12 R_0 + R + \frac{1}{6{m_0}^2}R^2
           - \frac{1}{{m_2}^2}\left(R_{\mu\nu}R^{\mu\nu}
              -\tfrac13R^2\right) 
        \right. } \notag \\ & \qquad\qquad \left.
           + \gamma\left(R_{\lm\mu\nu\rh}R^{\lm\mu\nu\rh}
              -4R_{\mu\nu}R^{\mu\nu}+R^2\right) \right].
\label{Riemannexpand}
\end{align}
Here we evaluate $f$ and its derivatives at the constant curvature 
solution, defining, for the contraction of the derivatives with a 
tensor $\Delta_{\lm\mu\nu\rh}$ which has the symmetries of the 
Riemann tensor but in otherwise arbitrary, 
\begin{equation}
\begin{aligned}
   \left. f \right|_{R_0} &= a_0, \\
   \left.\frac{df}{dR_{\lm\mu\nu\rh}}\right|_{R_0} \Delta_{\lm\mu\nu\rh}  
        &= a_1 \Delta, \\
   \left.\frac{d^2f}{dR_{\eta\kappa\lm\mu}dR_{\nu\rh\sg\tau}}\right|_{R_0} 
        \Delta_{\eta\kappa\lm\mu}\Delta_{\nu\rh\sg\tau} &= 
        a_{2,1} \Delta^2 + a_{2,2} \Delta_{\mu\nu}\Delta^{\mu\nu}
        + a_{2,3} \Delta_{\lm\mu\nu\rh}\Delta^{\lm\mu\nu\rh}, 
\end{aligned}
\end{equation}
with $\Delta_{\mu\nu}=g^{\lm\rh}\Delta_{\mu\lm\nu\rh}$ and 
$\Delta=g^{\mu\nu}\Delta_{\mu\nu}$. We have also introduced the parameters
\begin{equation}
\begin{aligned}
\beta &= a_1 - \left( a_{2,1} + \tfrac14 a_{2,2} 
              + \frac16 a_{2,3} \right) R_0, \\
{m_0}^2 &= \frac{\beta}{3a_{2,1}+a_{2,2}+a_{2,3}},
\end{aligned}
\qquad
\begin{aligned}
\gamma = \frac{a_{2,3}}{2\beta}, \\
{m_2}^2 = -\frac{2\beta}{a_{2,2}+4a_{2,3}},
\end{aligned}
\end{equation}
and have used the constant curvature condition $a_1R_0=2a_0$ in the final line 
of 
\eqref{Riemannexpand}. 

The final expression in \eqref{Riemannexpand} is a quadratic action with 
a Gauss-Bonnet term, a Weyl-squared term and a Ricci-scalar-squared term, 
together with a cosmological constant. The Gauss-Bonnet term can be dropped 
classically as a total divergence, leaving the action in the same quadratic 
form as discussed in our previous paper \cite{PRD-53-5583}. Reducing the action 
to 
a second-order form then follows 
exactly as in the case for $f(R_{\mu\nu})$ actions, so that we obtain the 
same transformed action \eqref{transfRicci}. Again we can verify that the 
transformed action has a vacuum solution at $R=R_0$ as is required for the 
consistency of our expansion. We also note that here too the expansion could 
have been made in terms of the variable $X_{\lm\mu\nu\rh}$, which could 
then be decomposed into its scalar and spin-two parts, the form of the 
decomposition depending on which vacuum is being considered.

In conclusion, general actions of the form $f(R_{\lm\mu\nu\rh})$ may have a 
variety of vacuum solutions, generically not apparent in a linear analysis. 
Around any vacuum the new degrees of freedom in the theory, aside from the 
massless graviton, can always be separated into a scalar field and a spin-two 
field. Unfortunately, the spin-two field is always ghost-like. 
  
\section{Higher-order Actions}
\label{fboxR}

In this section, we shall briefly discuss how to extend our analysis to 
actions with higher-order equations of motion. We have seen that 
an action involving any function of the curvature tensors gives at most 
fourth-order equations of motion. For higher-order equations we need to 
consider actions which include some derivatives of the curvature. Here, 
we shall be concerned with only the simplest form,
\begin{equation}
\label{boxRS}
   S = \frac{1}{2\kappa^2} \int{ d^4x \sqrt{-g} 
              f(R,\nabla^2R,\nabla^4R,\ldots,\nabla^{2k}R) },
\end{equation}
where we require that the function $f$ has been reduced, by integration by 
parts and dropping total derivatives, so as to minimize $k$. The
equations of motion following from similar actions have been considered
by Schmidt~\cite{CQG-7-1023} and Wands~\cite{CQG-11-269}, who showed them
to be equivalent to those for a set of scalar fields coupled to
gravity. Here will shall show the full equivalence in a new way, working
at the level of the action . 

It will be important to distinguish between possible forms 
of the function $f$. If we write $f=f(\lm_1,\lm_2,\ldots,\lm_{k+1})$, 
(where in the action we have $\lm_1=R$, $\lm_2=\nabla^2R$ and so on), 
we find there are two possible cases: either $\partial f/\partial\lm_{k+1}$ 
is a function of $\lm_{k+1}$ (case one) or it is not (case two). If it is not 
then it must be a function of $\lm_k$, since otherwise the original form was 
reducible; that is, the action was not written in a form minimized with 
respect to $k$. Thus, in case two, we can always decompose $f$ as
\begin{equation}
   \text{case 2:} \qquad f(\lm_1,\lm_2,\ldots,\lm_k,\lm_{k+1})
        = g(\lm_1,\lm_2,\ldots,\lm_k) \lm_{k+1} + h(\lm_1,\lm_2,\ldots,\lm_k). 
\end{equation}
Further, we see that for case one the equations of motion are 
$(4k+4)$th-order, while for case two they are $(4k+2)$th-order. 

We would like to reduce the action \eqref{boxRS} to a canonical second-order 
form, by introducing auxiliary fields. Given the order of the 
higher-derivative equations of motion we see that in case one we must 
introduce $2k+1$ new fields, while in case two we need only $2k$ new fields. 
The procedure we shall use is essentially a generalization of 
Ostrogradski's method for reducing a higher-order action to a first-order 
form \cite{MASP-VI4-385,Whittaker37}, only that here we shall be reducing to a 
second-order form. (The Ostrogradski result is usual given as a Hamiltonian, 
but this can always be rewritten as a Helmhotz Lagrangian, the analog of 
the form we shall use.)

We start by introducing a set of Lagrange multipliers, so 
\begin{multline}
   S = \frac{1}{2\kappa^2} \int{ d^4x } \sqrt{-g} \left[ 
          f(\lm_1,\lm_2,\ldots,\lm_{k+1}) + \mu\left(R-\lm_1\right) 
          \right. \\ \left.
          + \mu_1\left(\nabla^2\lm_1-\lm_2\right) + \cdots 
          + \mu_k\left(\nabla^2\lm_k-\lm_{k+1}\right) \right].
\end{multline}
Clearly eliminating fields via the Lagrange multiplier equations of motion, 
starting with $\mu_k$ and working down in $k$ returns one to the original 
action. 

However, we have introduced at least one too many new fields. It is clear
that the $\lm_{k+1}$ equation of motion is purely 
algebraic and eliminating it will not introduce higher-derivatives; that is, 
the action will remain second order. The $\lambda_{k+1}$ equation of motion 
reads
\begin{equation}
   \mu_k = 
       \begin{cases}
           {\displaystyle \frac{\partial f}{\partial\lm_{k+1}}
                (\lm_1,\lm_2,\ldots,\lm_k,\lm_{k+1})} & \text{in case 1}, \\
           {\displaystyle g(\lm_1,\lm_2,\ldots,\lm_k)} & \text{in case 2},
       \end{cases}
\end{equation}
where we have distinguished between the two cases discussed above, and 
substituted the special form of $f$ in case two. In case one, to form the 
analog of Ostrogradski's Lagrangian, we solve the equation to give 
$\lm_{k+1}$ as a function of $\lm_1,\dots,\lm_k$ and $\mu_k$, writing 
$\lm_{k+1}=\tlmk(\lm_1,\ldots,\lm_k,\mu_k)$. It should be noted that in 
general the solution is not unique, and we must divide the original theory 
into 
pieces corresponding to different branches of the solution, just as in the 
case of actions of the form $f(R)$ discussed in the previous section. In case 
two we simply substitute for $\mu_k$ and the special form of $f$. We get 
in case one
\begin{multline}
   \text{case 1:} \qquad S = \frac{1}{2\kappa^2} \int{ d^4x } \sqrt{-g} \left[ 
          f(\lm_1,\lm_2,\ldots,\lm_k,\tlmk(\lm_1,\ldots,\lm_k,\mu_k)) 
          + \mu\left(R-\lm_1\right)
          \right. \\ \left.
          + \mu_1\left(\nabla^2\lm_1-\lm_2\right) + \cdots 
          + \mu_k\left(\nabla^2\lm_k-\tlmk(\lm_1,\ldots,\lm_k,\mu_k)\right) 
          \right], 
\end{multline}
the exact analog of the Ostrogradski-Helmhotz Lagrangian, while for case two 
we have a slightly different form,
\begin{multline}
   \text{case 2:} \qquad S = \frac{1}{2\kappa^2} \int{ d^4x } \sqrt{-g} \left[ 
          h(\lm_1,\lm_2,\ldots,\lm_k) + \mu\left(R-\lm_1\right)
          \right. \\ \left.
          + \mu_1\left(\nabla^2\lm_1-\lm_2\right) + \cdots 
          + g(\lm_1,\lm_2,\ldots,\lm_k) \nabla^2\lm_k 
          \right]. 
\end{multline}
We see that, as expected, in case one we have a total of $2k+1$ auxiliary 
fields, while in case two we have only $2k$ new fields.

All that remains is to transform the action into a canonical form, growing 
canonical kinetic energy terms for all the new auxiliary fields. Let us 
concentrate on actions of case one. Terms of the form $\mu_i\nabla^2\lm_i$ 
are easy to deal with. Simply introducing a pair of new fields, $\chi_i$ 
and $\psi_i$ by
\begin{equation}
   \lm_i = \chi_i + \psi_i, \qquad \mu_i = \chi_i - \psi_i,
\end{equation}
we have, in the action,
\begin{align}
   \int{ d^4x\sqrt{-g} \mu_i\nabla^2\lm_i } &= \int{ d^4x\sqrt{-g} 
              \left[ \chi_i\nabla^2\chi_i - \psi_i\nabla^2\psi_i \right] }
              \notag \\
      &= \int{ d^4x\sqrt{-g} \left[
              - \left(\nabla\chi_i\right)^2 + \left(\nabla\psi_i\right)^2
              \right] }. 
\end{align}
We see that the $\chi_i$ field has a canonical kinetic term, but the $\psi_i$ 
field has the wrong sign; it is a ghost. This is characteristic of higher-
order 
theories. In reducing the theory to second-order the new fields always enter 
as a pair of a ghost-like field with a ordinary field. If we define a 
potential function,
\begin{multline}
   V(\chi_1,\ldots,\chi_k;\psi_1,\ldots,\psi_k) = 
           \mu\lm_1 + \mu_1\lm_2 + \cdots + \mu_{k-1}\lm_k 
           + \mu_k \tlmk(\lm_1,\ldots,\lm_k,\mu_k) \\
           - f(\lm_1,\lm_2,\ldots,\lm_k,\tlmk(\lm_1,\ldots,\lm_k,\mu_k))
\end{multline}
where it is understood that the right-hand side is evaluated at 
$\lm_i=\chi_i+\psi_i$, $\mu_i=\chi_i-\psi_i$, the action in case one 
becomes
\begin{equation}
   \text{case 1:} \qquad S = \frac{1}{2\kappa^2} \int{ d^4x \sqrt{-g} \left[ 
           \mu R - \sum_i \left\{ \left(\nabla\chi_i\right)^2 
                    - \left(\nabla\psi_i\right)^2 \right\}
           - V(\chi_1,\ldots,\chi_k;\psi_1,\ldots,\psi_k)  \right] }.
\end{equation}
To complete the transformation to canonical form all that is left is to make 
a conformal rescaling of the metric to remove the $\mu R$ coupling. As usual 
we define $\chi=\log\mu$ and rescale to a metric 
$\bg_{\mu\nu}=e^\chi g_{\mu\nu}$, giving 
\begin{multline}
\label{canonboxR1}
   \text{case 1:} \qquad S = \frac{1}{2\kappa^2} \int{d^4x} \sqrt{-\bg} \left[ 
           \bR - \tfrac32 \left(\bnabla\chi\right)^2
           - e^{-\chi} \sum_i \left\{ \left(\bnabla\chi_i\right)^2 
                    - \left(\bnabla\psi_i\right)^2 \right\}
           \right. \\ \left.
           - e^{-2\chi} V(\chi_1,\ldots,\chi_k;\psi_1,\ldots,\psi_k)  \right] 
.
\end{multline}
We conclude that the original case one higher-derivative gravity theory is
equivalent to canonical Einstein gravity coupled to $2k+1$ scalar fields,
$k+1$ of which, $\chi$ and $\chi_i$, propagate physically and $k$ of which,
$\psi_i$, are ghost-like.

To put the case two action in canonical form is more complicated because of 
the \linebreak 
$g(\lm_1,\lm_2,\ldots,\lm_k)\nabla^2\lm_k$ term. However, in principle, 
it is always possible to introduce a set of new fields 
$\{\chi_1,\ldots,\chi_k;\psi_1,\ldots,\psi_k\}$ which simultaneously 
diagonalize the kinetic terms for the $\lm_i$ and $\mu_i$, though now the 
form of the transformation will depend on the function $g$. At least $k-1$ of 
the new fields will be ghost-like. We can then make a conformal 
rescaling as in case one to put the action in the same canonical form 
\eqref{canonboxR1}, though the potential function will have a different form.

The conclusion is that there is a procedure for rewriting the higher-order 
action \eqref{boxRS} in a canonical second-order form. However, $k$ of the 
new fields in case one and at least $k-1$ of the fields in case two will 
be ghost-like. In each case we obtain a specific potential, and so we can 
again look for vacuum states as stationary points of the potential. 
Generically, as in the case of $f(R)$ actions 
discussed in the previous section, the non-trivial vacua do not correspond to 
the flat-space solution of the original higher-derivative theory. 

As an simple example of this procedure consider the function 
$f=R+\alpha\left(\nabla^2R\right)^2$. This example is of the first case since 
writing $f(\lm_1,\lm_2)=\lm_1+\alpha{\lm_2}^2$ with $\lm_1=R$ and 
$\lm_2=\nabla^2R$, we have $\partial f/\partial\lm_2=2\alpha\lm_2$, which 
is not independent of $\lm_2$. Following our general procedure, the first 
step is to introduce a set of Lagrange multipliers,  
\begin{align}
   S &= \frac{1}{2\kappa^2} \int{ d^4x \sqrt{-g} \left[ 
         R + \alpha\left(\nabla^2R\right)^2 \right] } \notag \\
     &= \frac{1}{2\kappa^2} \int{ d^4x \sqrt{-g} 
         \left[ \lm_1 + \alpha{\lm_2}^2 
         + \mu \left(R-\lm_1\right) + \mu_1 \left(\nabla^2\lm_1-\lm_2\right)
         \right] }. 
\end{align}
As discussed above we have introduced one too many auxiliary fields. We can 
eliminate $\lm_2$ by solving its equation of motion, which reads
\begin{equation}
   2\alpha\lm_2 - \mu_1 = 0
\end{equation}
implying $\lm_2=\mu_1/2\alpha$. Substituting back into the action gives
\begin{equation}
   S = \frac{1}{2\kappa^2} \int{ d^4x \sqrt{-g} \left[ \mu R 
         + \mu_1\nabla^2\lm_1 + \lm_1 - \mu\lm_1 - \tfrac14\alpha{\mu_1}^2
         \right] }. 
\end{equation}
Next, to put the kinetic energy for $\lm_1$ and $\mu_1$ in canonical form, we 
define $\lm_1=\chi_1+\psi_1$ and $\mu_1=\chi_1-\psi_1$, so that 
\begin{equation}
   S = \frac{1}{2\kappa^2} \int{ d^4x \sqrt{-g} \left[ \mu R 
         - \left(\nabla\chi_1\right)^2 + \left(\nabla\psi_1\right)^2 
         - \left(\mu-1\right)\left(\chi_1+\psi_1\right) 
         - \tfrac14\alpha\left(\chi_1-\psi_1\right)^2 
         \right] }. 
\end{equation}
Finally we make the conformal rescaling $\bg_{\mu\nu}=e^\chi g_{\mu\nu}$ with 
$\chi=\log\mu$ to put the action in canonical form
\begin{multline}
   S = \frac{1}{2\kappa^2} \int{d^4x} \sqrt{-\bg} \left[ \bR 
         - \tfrac32 \left(\bnabla\chi\right)^2
         - e^{-\chi} \left(\bnabla\chi_1\right)^2 
         + e^{-\chi} \left(\bnabla\psi_1\right)^2
         \right. \\ \left. 
         - e^{-2\chi} \left\{ \left(e^\chi-1\right)\left(\chi_1+\psi_1\right) 
                  + \tfrac14\alpha\left(\chi_1-\psi_1\right)^2 \right\} 
         \right]. 
\end{multline}
Thus we see that the original higher-derivative gravity theory is equivalent 
to canonical Einstein gravity coupled to three scalar fields. One field is 
ghost-like, and the potential has a single, unstable stationary point at 
$\chi=\chi_1=\psi_1=0$.

As a second example consider $f=\alpha+\beta R+\gamma R^2+\epsilon R\nabla^2R$
Writing  $\lm_1=R$ and $\lm_2=\nabla^2R$, we now have 
$\partial f/\partial\lm_2=\epsilon\lm_1$, independent of $\lm_2$, so this 
example clearly falls under case two. Repeating our procedure, we introduce 
Lagrange multipliers to give
\begin{align}
   S &= \frac{1}{2\kappa^2} \int{ d^4x \sqrt{-g} \left[ 
         \alpha + \beta R + \gamma R^2 + \epsilon R\nabla^2R \right] }
         \notag \\
     &= \frac{1}{2\kappa^2} \int{ d^4x \sqrt{-g} \left[ 
         \alpha + \beta\lm_1 + \gamma{\lm_1}^2 
         + \epsilon\lm_1\lm_2 + \mu \left(R-\lm_1\right) 
         + \mu_1 \left(\nabla^2\lm_1-\lm_2\right) \right] }. 
\end{align}
Again we have too many new fields. The $\lm_2$ equation of motion now reads 
\begin{equation}
   \epsilon\lm_1 = \mu_1,
\end{equation}
which cannot be solved for $\lm_2$ since the action is case two. However 
substituting this solution into the action eliminates both $\lm_2$ and 
$\mu_1$,  
leaving the second-order form, 
\begin{equation}
   S = \frac{1}{2\kappa^2} \int{ d^4x \sqrt{-g} \left[ 
         \mu R - \epsilon\ (\nabla\lm_1)^2 - \mu\lm_1 
             + \alpha + \beta\lm_1 + \gamma{\lm_1}^2 
         \right] }.
\end{equation}
The kinetic energy for $\lm_1$ is already in canonical form, but we must make 
a final conformal rescaling by $e^\chi=\mu$ to put the action in the 
completely canonical form
\begin{equation}
   S = \frac{1}{2\kappa^2} \int{ d^4x \sqrt{-\bg} \left[ \bR 
         - \tfrac32 \left(\bnabla\chi\right)^2
         - \epsilon e^{-\chi} \left(\bnabla\lm_1\right)^2 - V(\lm_1,\chi) 
         \right] }, 
\end{equation}
where we have the potential, 
\begin{equation}
   V(\lm_1,\chi) = e^{-2\chi} \left( e^\chi\lm_1 - \alpha -\beta\lm_1 
                            - \gamma{\lm_1}^2 \right)
\end{equation}
Again the original higher-derivative gravity is shown to be equivalent to 
ordinary Einstein gravity though now coupled to only two scalar fields. By 
choosing $\epsilon>0$ we can ensure that neither field is ghost-like. Note 
that we argued above that case two theories must have at least $k-1$ 
ghost-like fields. It is thus only in this special case of $k=1$ that we are 
able to have all the scalar fields non-ghost-like.

As before we obtain a specific potential for the fields. We find that 
$V(\lm_1,\chi)$ has a single stationary point at $\lm_1=-2\alpha/\beta$, 
$\chi=\ln(\beta-4\alpha\gamma/\beta)$, provided $\beta-4\alpha\gamma/\beta>0$. 
Expanding around this point we find that, if $\alpha<0$ and $\beta<0$, we have
a stable minimum. The value of the potential at the minimum is 
$V(\lm_1,\chi)=-\alpha/\left(\beta^2-4\alpha\gamma\right)$ which is negative. 
We conclude that, for the given range of $\alpha$, $\beta$ and $\gamma$, the 
theory has a single stable vacuum state with negative 
cosmological constant. From Einstein's equation we find that this state is an 
anti-deSitter space with $\bR=-2\alpha/\left(\beta^2-4\alpha\gamma\right)$. 
Using the fact that $R=e^\chi\bR$ for covariantly constant $\chi$, we find 
that 
$R=-2\alpha/\beta$. Thus the vacuum state also corresponds to an anti-deSitter 
space in the original higher-derivative theory. 

\section{Conclusion}
\label{conclusion}

The most important conclusion of this paper is that higher-derivative 
theories of gravitation generically have multiple stable vacua. One of these 
may be trivial, corresponding to flat spacetime, but all the other vacua are 
non-trivial with the associated manifold being either deSitter or 
anti-deSitter spacetime with non-vanishing cosmological constant. While of 
interest from various points of view, such non-trivial vacua cannot represent 
the universe as it is now since the radius of curvature of these solutions 
is of order the inverse Planck mass. Thus, one might conclude that non-trivial 
gravitational vacua are irrelevant for particle physics theories. 
However, this is not the case. We have recently shown that if we extend the 
methods and results of this paper to the realm of $N=1$ supergravity, then 
non-trivial vacua can exist with vanishing cosmological constant 
\cite{NPB-476-175}. Furthermore, we find that supersymmetry is generically 
spontaneously broken in these vacuum states. It follows that higher-derivative 
$N=1$ supergravitation could play a pivotal role in high energy physics. This 
possibility is being pursued elsewhere \cite{PLB-381-154}.

\section*{Acknowledgments}

This work was supported in part by DOE Grant No.\ DE-FG02-95ER40893 and 
NATO Grand No.\ CRG-940784.

\end{document}